\documentclass[letter]{emulateapj}

\usepackage{amsmath}
\usepackage{comment}

\shorttitle{Astrometry in Wide-Field Surveys}
\shortauthors{P\'al \& Bakos}

\newcommand{\ctbd}[1]{}

\newcommand{\ccdsize}[1]{\ensuremath{\rm #1\times\rm#1}}

\newcommand{\ordo}{\ensuremath{\mathcal{O}}}

\newcommand{\pack}[1]{\textsc{\lowercase{#1}}}
\newcommand{\prog}[1]{\texttt{\lowercase{#1}}}

\newcommand{\fihat}{\pack{fihat}}

\newcommand{\figr}[1]{Fig.~\ref{fig:#1}}
\newcommand{\secr}[1]{\mbox{\S\ \ref{sec:#1}}}
\newcommand{\eqr}[1]{Eq.~\ref{eq:#1}}

\newcommand{\aref}{\ensuremath{\mathcal{R}}}
\newcommand{\areft}{\ensuremath{\mathcal{R^{\prime}}}}
\newcommand{\aimg}{\ensuremath{\mathcal{I}}}
\newcommand{\atran}{\ensuremath{\mathcal{F}_{R\rightarrow I}}}
\newcommand{\atranone}{\ensuremath{\mathcal{F}^{(1)}_{R\rightarrow I}}}
\newcommand{\atrani}{\ensuremath{\mathcal{F}^{(i)}_{R\rightarrow I}}}
\newcommand{\tmixed}{\ensuremath{T^{\text{(mix)}}}}
\newcommand{\tchir}{\ensuremath{T^{\text{(chir)}}}}
\newcommand{\tcont}{\ensuremath{T^{\text{(cont)}}}}

\begin{document}
\title{Astrometry in Wide-field Surveys}
\author{Andr\'as P\'al\altaffilmark{1}}
\affil{Lor\'and E\"otv\"os University, Department of Astronomy, Budapest, 
Hungary, H-1117}
\email{apal@cfa.harvard.edu}

\and
\author{G\'asp\'ar \'A.~Bakos\altaffilmark{2}}
\affil{Harvard-Smithsonian Center for Astrophysics, Cambridge, MA 02138}

\altaffiltext{1}{Visiting Astronomer at the Harvard-Smithsonian Center for 
Astrophysics}
\altaffiltext{2}{Hubble Fellow}

\begin{abstract}

We present a robust and fast algorithm for performing astrometry and
source cross-identification on two dimensional point lists, such as
between a catalogue and an astronomical image, or between two images.
The method is based on minimal assumptions: the lists can be rotated,
magnified and inverted with respect to each other in an arbitrary way. 
The algorithm is tailored to work efficiently on wide fields with large
number of sources and significant non-linear distortions, as long as
the distortions can be approximated with linear transformations
locally, over the scale-length of the average distance between the
points.  The procedure is based on symmetric point matching in a newly
defined continuous triangle space that consists of triangles generated
by an extended Delaunay triangulation. Our software implementation
performed at the 99.995\% success rate on $\sim260,000$ frames taken by
the HATNet project.

\end{abstract}



\section{Introduction}
\label{sec:intro}

Cross-matching two two-dimensional points lists is a crucial step in
astrometry and source identification. The tasks involves finding the
appropriate geometrical transformation that transforms one list into
the reference frame of the other, followed by finding the best matching
point-pairs.
One of the lists usually contains the pixel coordinates of sources in
an astronomical image (e.g.~point-like sources, such as stars), while
the other list can be either a reference catalog with celestial
coordinates, or it can also consist of pixel coordinates that originate
from a different source of observation (another image). Throughout this
paper we denote the reference (list) as \aref, the image (list) as
\aimg, and the function that transforms the reference to the image as
\atran.

The difficulty of the problem is that in order to find matching pairs,
one needs to know the transformation, and vica versa: to derive the
transformation, one needs point-pairs. Furthermore, the lists may not
fully overlap in space, and may have only a small fraction of sources
in common.

By making simple assumptions on the properties of \atran, however, the
problem can be tackled. A very specific case is when there is
only a simple translation between the lists, and one can use
cross-correlation techniques \citep[see][]{phillips95} to find the
transformation. 
We note, that the a method proposed by \citet{thiebaut01} uses the
whole image information to derive a transformation (translation and
magnification).

A more general assumption, typical to astronomical applications, is a
that $\atran$ is a similarity transformation (rotation, magnification,
inversion, without shear), i.e.~$\atran = \lambda
\mathbf{A}\underline{r} + \underline{b}$, where $\mathbf{A}$ is a
(non-zero) scalar $\lambda$ times the orthogonal matrix,
$\underline{b}$ is an arbitrary translation, and $\underline{r}$ is the
spatial vector of points. Exploiting that geometrical patterns remain
similar after the transformation, more general algorithms have been
developed that are based on pattern matching \citep{groth86,valdes95}.
The idea is that
the initial transformation is found by the aid of a specific set of
patterns that are generated from a subset of the points on both \aref\
and \aimg. For example, the subset can be that of the brightest
sources, and the patterns can be triangles.  With the knowledge of this
initial transformation, more points can be cross-matched, and the
transformation between the lists can be iteratively refined.
Some of these methods are implemented as an
\pack{IRAF}
task\footnote{IRAF is distributed by the National Optical Astronomy
Observatories, which are operated by the Association of Universities
for Research in Astronomy, Inc., under cooperative agreement with the
National Science Foundation.}
in \prog{immatch} \citep{phillips95}.

The above pattern matching methods perform well as long as the dominant
term in the transformation is linear, such as for astrometry of narrow
field-of-view (FOV) images, and as long as the number of sources is
small (because of the large number of patterns that can be generated
\--- see later).  In the past decade of astronomy, with the development of
large format CCD cameras or mosaic imagers, many wide-field surveys
appeared, such as those looking for transient events 
(e.g.~ROTSE --- \citealt{akerlof00}), 
transiting planets (e.g.~Kelt -- \citealt{pepper04},
TrES \--- \citealt{alonso04},
HATNet \--- \citealt{bakos02,bakos04},
see \citealt{dc06} for further references), 
or all-sky variability (e.g.~ASAS \--- \citealt{pojmanski97}). There
are non-negligible, higher order distortion terms in the astrometric
solution that are due to, for instance, the projection of celestial to
pixel coordinates and the properties of the fast focal ratio optical
systems. Furthermore, these images may contain $\sim 10^5$ sources, and
pattern matching is non-trivial.

These surveys necessitated a further generalization of the algorithm,
which we present in this paper. To be more specific, we were motivated
by the astrometric requirements of the Hungarian-made Automated
Telescope Network (HATNet). Each HAT telescope of the Network consists
of a $f=200\mathrm{mm}$, $d=f/1.8$ telephoto lens and a \ccdsize{2K}
CCD yielding an $8\arcdeg\times8\arcdeg$ FOV. In our experience, we
need at least 4th order polynomial functions of the pixel coordinates
in order to properly describe the distortion of the lens. With a
typical exposure time of 5 minutes in I-band, in a moderately dense
field ($b \approx 15^\circ$) there are 30000 stars brighter than I=13
for which better than 10\% photometry can be achieved. If we consider
all 3-$\sigma$ detections, we have to deal with the identification of
$\sim$100,000 sources.

The algorithm presented in this paper is based on, and is a
generalization of the above pattern matching algorithms. It is very
fast, and works robustly for wide-field imaging with minimal
assumptions. Namely, we assume that:
i) the distortions are non-negligible, but small compared to the linear
term,
ii) there exists a smooth transformation between the reference and
image points, 
iii) the point lists have a considerable number of sources in common,
and
iv) the transformation is locally invertible.
The paper is presented as follows. First we describe symmetrical point
matching in \secr{pmatch} before we go on to the discussion of finding
the transformation (\secr{tran}). The software implementation and its
performance on a large and inhomogeneous dataset is demonstrated in
\secr{exam}. Finally, we draw conclusions in \secr{sum}.

\section{Symmetric point matching}
\label{sec:pmatch}

First, let us assume that \atran\ is known. To find point-pairs between
\aref\ and \aimg\, one should first transform the reference points to
the reference frame of the image: $\areft = \atran(\aref)$. Now it is
possible to perform a simple symmetric point matching between \areft\
and \aimg. One point ($R_1\in\areft$) from the first and one point
($I_1\in\aimg$) from the second set are treated as a pair if the
closest point to $R_1$ is $I_1$ \emph{and} the closest point to
$I_1$ is $R_1$. This requirement is symmetric by definition and excludes such
cases when e.g.~the closest point to $R_1$ is $I_1$, but there exists an
$R_2$ that is even closer to $I_1$, etc.

In one dimension, finding the point of a given list nearest to a
specific point ($x$) can be implemented as a binary search. Let us
assume that the point list with $N$ points is ordered in ascending
order.  This has to be done only once, at the beginning, and using the
quicksort algorithm, for example, the required time scales on average
as $\ordo(N \log N)$. Then $x$ is compared to the median of the list:
if it is less than the median, the search can be continued recursively
in the first $N/2$ points, if it is greater than the median, the second
$N/2$ half is used. At the end only one comparison is needed to find
out whether $x$ is closer to its left or right neighbor, so in total
$1+\log_2(N)$ comparisons are needed, which is an $\ordo(\log N)$
function of $N$. Thus, the total time including the initial sorting
also goes as $\ordo(N \log N)$.

As regards a two dimensional list, let us assume again, 
that the points are ordered in
ascending order by their $x$ coordinates (initial sorting $\sim \ordo(N
\log N)$), and they are spread uniformly in a square of unit area.
Finding the nearest point in $x$ coordinate also requires $\ordo(\log
N)$ comparisons, however, the point found presumably will not be the
nearest in Euclidean distance. The expectation value of the distance
between two points is $1/\sqrt{N}$, and thus we have to compare points
within a strip with this width and unity height, meaning
$\ordo(\sqrt{N})$ comparisons. Therefore, the total time required by a
symmetric point matching between two catalogs in two dimensions
requires $\ordo(N^{3/2} \log N)$ time.

We note that finding the closest point within a given set of points is
also known as nearest neighbor problem \citep[for a summary see][and
references therein]{gionis02}.  It is possible to reduce the
computation time in 2 dimensions to $\ordo(N \log N)$ by the aid of
Voronoi diagrams and Voronoi cells, but we have not implemented such an
algorithm in our matching codes.

\section{Finding the transformation}
\label{sec:tran}

Let us go back to finding the transformation between \aref\ and \aimg.
The first, and most crucial step of the algorithm is to find an initial
``guess'' \atranone\ for the transformation based on a variant of
triangle matching. Using \atranone, \aref\ is transformed to
\aimg, symmetric point-matching is done, and the paired coordinates are
used to further refine the transformation (leading to \atrani\ in iteration
$i$), and increase the number of matched points iteratively. A major part of
this paper is devoted to finding the initial transformation.

\subsection{Triangle matching}
\label{sec:trimatch}

It was proposed earlier by \citet{groth86} and \citet{stetson89}, and
recently by others \citep[see][]{valdes95} to use triangle matching for
the initial ``guess'' of the transformation. 
The total number of triangles that can be formed using $N$ points is
$N(N-1)(N-2)/6$, an $\ordo(N^3)$ function of $N$. As this can
be an overwhelming number, one can resort to using a subset of the points
for the vertices of the triangles to be generated. One can also limit
the parameters of the triangles, such as exclude elongated or large
(small) triangles.

As triangles are uniquely defined by three parameters, for example the
length of the three sides, these parameters (or their appropriate
combinations) naturally span a 3-dimensional triangle space. Because
our assumption is that \atran\ is dominated by the linear term, to
first order approximation there is a single scalar magnification
between \aref\ and \aimg\ (besides the rotation, chirality and
translation). It is possible to reduce the triangle space to a
normalized, two-dimensional triangle space ($(T_x,T_y)\in T$), whereby
the original size information is lost. Similar triangles (with or
without taking into account a possible flip) can be represented by the
same points in this space, alleviating triangle matching between \aref\
and \aimg.

\subsubsection{Triangle spaces}
\label{sec:trispace}

\begin{figure}
\plotone{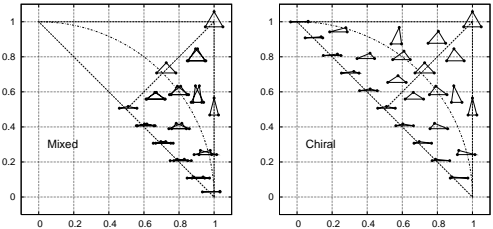}
\caption{
The position of triangles in the mixed and the chiral triangle spaces.
The exact position of a given triangle is represented by its center of
gravity.  Note that in the mixed triangle space some triangles with
identical side ratios but different orientation overlap. The dashed
line shows the boundaries of the triangle space. The dotted-dashed line
represents the right triangles and separates obtuse and acute ones.
\label{fig:trichir}}
\end{figure}

There are multiple ways of deriving normalized triangle spaces. One can
define a ``mixed'' normalized triangle space \tmixed, where the
coordinates are insensitive to inversion between the original
coordinate lists, i.e.~all similar triangles are represented by the
same point irrespective of their chirality \citep{valdes95}:
\begin{eqnarray}
\tmixed_x & = & p/a, \label{eq:tmix1}\\
\tmixed_y & = & q/a, \label{eq:tmix2}
\end{eqnarray}
where $a$, $p$ and $q$ are the sides of the triangle in descending
order. Triangles in this space are shown on the left panel of
\figr{trichir}.
Coordinates in the mixed triangle space are continuous functions of the
sides (and therefore of the spatial coordinates of the vertices of the
original triangle) but the orientation information is lost.  Because we
assumed that \atran\ is smooth and bijective, no local inversions and
flips can occur. In other words, \aref\ and \aimg\ are either flipped
or not with respect to each other, but chirality does not have a
spatial dependence, and there are no ``local spots'' that are mirrored.
Therefore, using mixed triangle space coordinates can yield false
triangle matchings that can lead to an inaccurate initial
transformation, or the match may even fail. Thus, for large sets of
points and triangles it is more reliable to fix the orientation of the
transformation. For example, first assume the coordinates are not
flipped, perform a triangle match, and if this match is unsatisfactory,
then repeat the fit with flipped triangles.

This leads to the definition of an alternative, ``chiral'' triangle
space:
\begin{eqnarray}
\tchir_x & = & b/a, \label{eq:tchir1}\\
\tchir_y & = & c/a, \label{eq:tchir2}
\end{eqnarray}
where $a$, $b$ and $c$ are the sides in counter-clockwise order and $a$
is the longest side. In this space similar triangles with different
orientations have different coordinates. The shortcoming of \tchir\ is
that it is not continuous: a small perturbation of an isosceles
triangle can result in a new coordinate that is at the upper rightmost
edge of the triangle space.

In the following, we show that it is possible to define a
parametrization that is both continuous and preserves chirality. Flip
the chiral triangle space in the right panel of \figr{trichir} along
the $T_x + T_y=1$ line. This transformation moves the equilateral
triangle into the origin. Following this, apply radial magnification of
the whole space to move the $T_x + T_y=1$ line to the $T_x^2 + T_y^2=1$
arc (the magnification factor is not constant: $1$ along the direction
of $x$ and $y$-axis and $\sqrt{2}$ along the $T_x=T_y$ line).  Finally,
apply an azimuthal slew by a factor of $4$ to identify the
$T_y=0, T_x>0$ and $T_x=0, T_y>0$ edges of the space. To be more
specific, let us denote the sides as in \tchir: $a$, $b$ and
$c$ in counter-clockwise order where $a$ is the longest, and define
\begin{eqnarray}
\alpha & = & 1-b/a, \\
\beta & = & 1-c/a.
\end{eqnarray}
Using these values, it is easy to prove that by using the definitions
of the following variables:
\begin{eqnarray}
x_1 & = & \frac{\alpha(\alpha+\beta)}{\sqrt{\alpha^2+\beta^2}}, \\
y_1 & = & \frac{\beta(\alpha+\beta)}{\sqrt{\alpha^2+\beta^2}}, \\
x_2 & = & x_1^2-y_1^2, \\
y_2 & = & 2x_1y_1, 
\end{eqnarray}
one can define the triangle space coordinates as:
\begin{eqnarray}
\tcont_x & = & \frac{x_2^2-y_2^2}{(\alpha+\beta)^3} = 
	\frac{(\alpha+\beta)\left(\alpha^4-6\alpha^2\beta^2+\beta^4\right)}
	{(\alpha^2+\beta^2)^2}, \label{eq:tcnt1}\\
\tcont_y & = & \frac{2x_2y_2}{(\alpha+\beta)^3} =
 	\frac{4(\alpha+\beta)\alpha\beta(\alpha^2-\beta^2)}
	{(\alpha^2+\beta^2)^2}\, .\label{eq:tcnt2}
\end{eqnarray}

\begin{figure}
\plotone{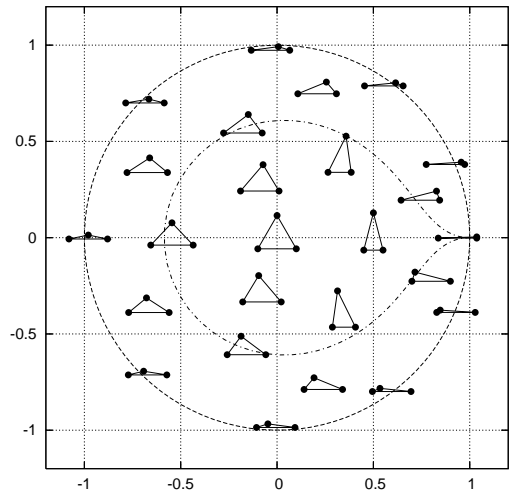}
\caption{
Triangles in the the continuous triangle space as defined by
Eqs.~\ref{eq:tcnt1}\---\ref{eq:tcnt2}. 
We show the same triangles as earlier, in \figr{trichir}, for the
\tmixed\ and \tchir triangle spaces. Equilateral triangles are centered
in the origin. The dotted-dashed line refers to the right triangles,
and divides the space to acute (inside) and obtuse (outside) triangles.
Isosceles triangles are placed on the $x$-axis (where
$\tcont_y=0$).
\label{fig:tricont}}
\end{figure}

The above defined \tcont\ continuous triangle space has many
advantages. It is a continuous function of the sides for all
non-singular triangles, and also preserves chirality information.
Furthermore, it spans a larger area, and misidentification of triangles
(that may be very densely packed) is decreased. Some triangles in this
space are shown in \figr{tricont}.

\notetoeditor{This is the original place of \figr{tricont}. This is a
two-panel figure. We would request placing this figure close to the
context, and in such a way that one panel width is roughly one column
in the paper.}

\subsubsection{Optimal triangle sets}
\label{sec:opttri}

As it was mentioned before, the total number of triangles that can be
formed from $N$ points is $\approx N^3/6$.  Wide-field images typically
contain $\ordo(10^4)$ points or more, and the total number of triangles
that can be generated -- a complete triangle list -- is unpractical for
the following reasons.
First, storing and handling such a large number of triangles with
typical computers is inconvenient. To give an example, a full
triangulation of 10,000 points yields $\sim 1.7\times10^{11}$ triangles.

Second, this complete triangle list includes many triangles that are
not optimal to use. For example large triangles can be significantly
distorted in \aimg\ with respect to \aref, and thus are represented by
substantially different coordinates in the triangle space. The size of
optimal triangles is governed by two factors: the distortion of large
triangles, and the uncertainty of triangle parameters for small
triangles that are comparable in size to the astrometric errors of the
vertices.

To make an estimate of the optimal size for triangles, let us denote
the characteristic size of the image by $D$, the astrometric error by
$\delta$, and the size of a selected triangle as $L$. For the sake of
simplicity, let us ignore the distortion effects of a complex optical
assembly, and estimate the distortion factor $f_d$ in a wide field
imager as the difference between the orthographic and gnomonic
projections
\citep[see][]{calabretta02}:
\begin{equation}
f_d \approx |(\sin(d)-\tan(d))/d| \approx |1-\cos(d)|\,,
\end{equation}
where $d$ is the radial distance as measured from the center of the
field. For the HATNet frames ($d = D\approx6\arcdeg$ to the corners)
this estimate yields $f_d\approx 0.005$.
The distortion effects yield an error of $f_d L/D$ in the triangle
space \--- the bigger the triangle, the more significant the
distortion. For the same triangle, astrometric errors cause an
uncertainty of $\delta/L$ in the triangle space that decreases with
increasing $L$. Making the two errors equal,
\begin{equation}
\frac{f_d \cdot L}{D} = \frac{\delta}{L},
\end{equation}
an optimal triangle size can be estimated by
\begin{equation}
L_{\text{opt}}=\sqrt{\frac{\delta \cdot D}{f_d}}.
\end{equation}
In our case $d=2048$ pixels (or $6\arcdeg$), $f_d=0.005$ and the
centroid uncertainty for an $I=11$ star is $\delta=0.01$, so the
optimal size of the triangles is $L_{\text{opt}}\approx 60-70$ pixels.

Third, dealing with many triangles may result in a triangle space that
is over-saturated by the large number of points, and may yield
unexpected matchings of triangles. In all definitions of the previous
subsection, the area of the triangle space is approximately unity.
Having triangles with an error of $\sigma$ in triangle space and
assuming them to have a uniform distribution, allowing a $3\sigma$
spacing between them, and assuming $\sigma=\delta/L_{\text{opt}}$, the
number of triangles is delimited to:
\begin{equation}
T_{\text{max}}\approx
\frac{1}{(3\sigma)^2}\approx
\frac{1}{9}\left(\frac{L}{\delta}\right)^2=
\frac{D}{9f_d\delta}\,.
\end{equation}
In our case (see values of $D$, $f_d$ and $\delta$ above) the former
equation yields $T_{\text{opt}}\approx 2\times 10^6$ triangles. Note
that this is 5 orders of magnitude smaller than a complete
triangulation ($\ordo(10^{11})$). 

\subsubsection{The extended Delaunay triangulation}
\label{sec:extdel}

\begin{figure}
\plotone{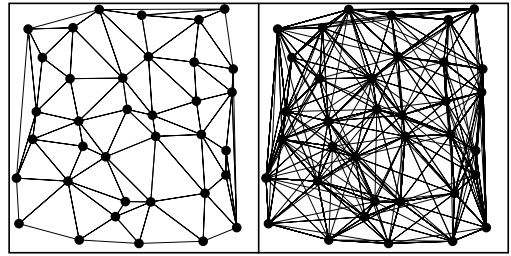}
\caption{
Triangulations of some randomly distributed points: the left panel
shows the Delaunay triangulation (60 triangles in total) 
the right panel exhibits the $\ell=1$ extended triangulation 
(312 triangles) of the same point set.
\label{fig:delaunay}}
\end{figure}

Delaunay triangulation \citep[see][]{shewchuk96} is a fast and robust
way of generating a triangle mesh on a point-set. The Delaunay
triangles are disjoint triangles where the circumcircle of any triangle
contains no other points from any other triangle. This is also
equivalent to the most efficient exclusion of distorted triangles in a
local triangulation. For a visual example of a Delaunay triangulation
of a random set of points, see the left panel of \figr{delaunay}.

\notetoeditor{This is the original place of \figr{delaunay}. This is a
two-panel figure. We would request placing this figure close to the
context, and in such a way that one panel width is roughly one column
in the paper.}

Following Euler's theorem (also known as the polyhedron formula), one
can calculate the number of triangles in a Delaunay triangulation of
$N$ points:
\begin{equation}
T_D = 2N-2-C,
\end{equation}
where $C$ is the number of edges on the convex hull of the point set.
For large values of $N$, $T_D$ can be estimated as $2N$, as $2+C$ is
negligible. Therefore, if we select a subset of points (from \aref\ or
\aimg) where neighboring ones have a distance of $L_{\text{opt}}$, we
get a Delaunay triangulation with approximately $2D^2/L_{\text{opt}}^2$
triangles. The $D$, $\delta$ and $f_d$ values for HAT images
correspond to $\approx 6000$ triangles, i.e.~3000 points. 
In our experience, this yields very fast matching, but it is not robust
enough for general use, because of the following reasons.

Delaunay triangulation is very sensitive for removing a point
from the star list.  According to the polyhedron formula, on the
average, each point has 6 neighboring points and belongs to 6
triangles. Because of observational effects or unexpected events, the
number of points fluctuates in the list. To mention a few examples, it
is customary to build up \aimg\ from the brightest stars in an image,
but stars may get saturated or fall on bad columns, and thus disappear
from the list. Star detection algorithms may find sources depending on
the changing full-width at half maximum (FWHM) of the frames.
Transients, variable stars or minor planets can lead to additional
sources on occasions. In general, if one point is removed, 6 Delaunay
triangles are destroyed and 4 new ones are formed that are totally
disjoint from the 6 original ones (and therefore they are represented
by substantially different points in the triangle space). Removing one
third of the generating points might completely change the
triangulation\footnote{Imagine a honey-bee cell structure where all
central points of the hexagons are added or removed: these two
construction generates disjoint Delaunay triangulations.}.

Second, and more important, there is no guarantee that the spatial {\em
density} of points in \aref\ and \aimg\ is similar. For example, the
reference catalog is retrieved for stars with different magnitude
limits than those found on the image. If the number of points in common
in \aref\ and \aimg\ is only a small fraction of the total number of
points, the triangulation on the reference and image has no common
triangles. 
Third, the number of the triangles  with Delaunay triangulation
($T_{\text{D}}$) is definitely smaller than $T_{\text{opt}}$; i.e.~the
triangle space could support more triangles without much confusion.

Therefore, it is beneficial to extend the Delaunay triangulation. A
natural way of extension can be made as follows. Define a level $\ell$
and for any given point ($P$) select all points from the point set of
$N$ points that
can be connected to $P$ via maximum $\ell$ edges of the Delaunay
triangulation. Following this, one can generate the \emph{full}
triangulation of this set and append the new triangles to the whole
triangle set. This procedure can be repeated for all points in the point
set at fixed $\ell$. For self-consistence, the $\ell=0$ case is
defined as the Delaunay triangulation itself. 
If all points have 6
neighbors, the number of ``extended'' triangles {\em per data point} is:
\begin{equation}
T_\ell=(3\ell^2+3\ell+1)(3\ell^2+3\ell)(3\ell^2+3\ell-1)/6
\end{equation}
for $\ell>0$, 
i.e.~this extension introduces $\ordo(\ell^6)$ new triangles. 
Because some of the extended triangles are repetitions of other
triangles from the original Delaunay triangulation and from the
extensions of another points, the final dependence only goes as
$\ordo(T_{\text{D}}\ell^2)$. We note that our software implementation
is slightly different, and the expansion requires $\ordo(N\ell^2)$ time
and automatically results in a triangle set where each triangle is
unique. To give an example, for $N=10,000$ points the Delaunay
triangulation gives $20,000$ triangles, the $\ell = 1$ extended
triangulation gives $\sim 115,000$ triangles, $\ell = 2$ some $\sim
347,000$ triangles, $\ell = 3$ $875,000$ and $\ell = 4$ $\sim
1,841,000$ triangles, respectively. The extended triangulation is not
only advantageous because of more triangles, and better chance for
matching, but also, there is a bigger variety in size that enhances
matching if the input and reference lists have different spatial
density.

\subsubsection{Matching the triangles in triangle space}
\label{sec:trimatch2}

If the triangle sets for both the reference and the input list are
known, the triangles can be matched in the normalized triangle space
(where they are represented by two dimensional points) using the
symmetric point matching as described in \secr{pmatch}.

In the next step we create a ${N_R\times N_I}$ ``vote'' matrix $V$,
where $N_R$ and $N_I$ are the number of points in the reference and
input lists that were used to generate the triangulations,
respectively. The elements of this matrix have an initial value of 0.
Each matched triangle corresponds to 3 points in the reference list
(identified by $r_1$, $r_2$, $r_3$) and 3 points in the input list
($i_1$, $i_2$ and $i_3$). Knowing these indices, the matrix elements
$V_{r_1i_1}$, $V_{r_2i_2}$ and $V_{r_3i_3}$ are incremented. The
magnitude of this increment (the \emph{vote}) can depend on the
distances of the matching triangles in the triangle space: the closer
they are, the higher votes these points get. In our implementation, if
$N_T$ triangles are matched in total, the closest pair gets $N_T$
votes, the second closest pair gets $N_T-1$ votes, and so on.

Having built up the vote matrix, we select the greatest elements of
this matrix, and the appropriate points referring to these row and
column indices are considered as matched sources. We note that not all
of the positive matrix elements are selected, because elements with
smaller votes are likely to be due to misidentifications. We found that
in practice the upper 40\% of the matrix elements yield a robust match.

\subsection{The unitarity of the transformations}
\label{sec:unit}

If an initial set of the possible point-pairs are known from
triangle-matching, one can fit a smooth function (e.g.~a polynomial)
that transforms the reference set to the input points.
Our assumption was that the dominant term in our transformation is the
similarity transformation, which implies that the homogeneous linear
part of it should be \emph{almost} almost a unitarity
operator.\footnote{Here 
$\mathbf{A}\mathbf{A^+} = I$, where $\mathbf{A^+}$
is the adjoint of $\mathbf{A}$ and $\mathbf{I}$ is the identity,
i.e.~$\mathbf{A}$ is an orthogonal transformation with possible
inversion and magnification.}
After the transformation is determined, it is useful to measure how much we
diverge from this assumption.
As mentioned earlier (\secr{intro}), similarity transformations
can be written as
\begin{equation}
\label{eq:lintr2}
\underline{r^{\prime}} = \lambda \mathbf{A} \underline{r} + \underline{b} 
\, \equiv \,
\lambda \begin{pmatrix}a&b\\c&d\end{pmatrix} \underline{r} +
\underline{b},\,
\end{equation}
where $\lambda\ne0$, and the $a,b,c,d$ matrix components are the sine
and cosine of a given rotational angle, i.e.~$a=c$ and $b=-c$.

If we separate the homogeneous linear part of the transformation, as
described by a matrix similar to that in \eqr{lintr2}, it will be a
combination of rotation and dilation with possible inversion if
$|a|\approx |d|$ and $|c|\approx |b|$.
We can define the unitarity of a matrix as:
\begin{equation}
\Lambda^2 := \frac{(a\mp d)^2+(b \pm c)^2}{a^2 + b^2 + c^2 + d^2}\,,
\end{equation}
where the $\pm$ indicates the definition for regular and inverting
transformations, respectively. For a combination of rotation and
dilation, $\Lambda$ is zero, for a distorted transformation
$\Lambda\approx f_d \ll 1$.

The $\Lambda$ unitarity gives a good measure of how well the initial
transformation was determined. It happens occasionally that the
transformation is erroneous, and in our experience, in these cases 
$\Lambda$ is not just larger than the expectational value of $f_d$, but
it is $\approx1$. This enables fine-tuning of the algorithm, such as
changing chirality of the triangle space, or adding further iterations
till satisfactory $\Lambda$ is reached. 

\subsection{Point matching in practice}
\label{sec:pract}

In practice, matching points between the \aref\ reference and \aimg\
image goes as the following:
\begin{enumerate}
\item Generate two triangle sets $T_R$ and $T_I$ on \aref\ and \aimg, 
respectively:
	\begin{enumerate}
	\item In the first iteration, generate only Delaunay triangles.
	\item Later, if necessary, extended triangulation can be generated 
		with increasing levels of $\ell$.
	\end{enumerate}
\item Match these two triangle sets in the triangle space using
symmetric point matching.
\item Select some possible point-pairs using a vote-algorithm (yielding 
$N_0$ pairs).
\item Derive the initial smooth transformation \atranone\ 
using a least-squares fit.
	\begin{enumerate}
	\item Check the unitarity of \atranone.
	\item If it is greater than a given threshold ($\ordo(f_d)$), 
		increase $\ell$ and go to step 1/b.
		If the unitarity is less than this threshold, proceed to step 5.
	\item If we reached the maximal allowed $\ell$, try the procedure
		with triangles that are flipped with respect to each other
		between the image and reference, i.e.~switch chirality of the
		\tcont\ triangle space. 
	\end{enumerate}

\item Transform \aref\ using this initial transformation 
to the reference frame of the image ($\areft = \atranone(\aref)$).

\item Perform a symmetric point matching between
\areft\ and \aimg\ (yielding $N_1 > N_0$ pairs).

\item Refine the transformation based on the greater number of 
pairs, yielding transformation \atrani, where $i$ is the iteration
number. 

\item If necessary, repeat points 5, 6 and 7 iteratively, increase the
number of matched points, and refine the transformation.
\end{enumerate}

For most astrometric transformations and distortions it holds that
locally they can be approximated with a similarity transformation.  At
a reasonable density of points on \aimg\ and \aimg, the triangles
generated by a (possibly extended) Delaunay triangulation are small
enough not to be affected by the distortions.  The crucial step is the
initial triangle matching, and due to the use of local triangles, it
proves to be robust procedure. It should be emphasized that \atrani\
can be any smooth transformation, for example an affine transformation
with small shear, or polynomial transformation of any reasonable order.
The optimal value of the order depends on the magnitude of the
distortion. The detailed description of fitting such models and
functions can be found in various textbooks
\citep[see e.g.~Chapter 15. in][]{press92}. It is noteworthy that in
step 7 one can perform a weighted fit with possible iterative rejection
of n-$\sigma$ outlier points.

\section{Software implementation and applications}
\label{sec:exam}

\subsection{Software implementation}
\label{sec:sw}

The coordinate matching and coordinate transforming algorithms are
implemented in two stand-alone binary programs written in ANSI C. The
program named \prog{grmatch} matches point sets, including triangle
space generation, triangle matching, symmetric point matching and
polynomial fitting, that is steps 1 through 4 in \secr{pract}. The
other program, \prog{grtrans}, transforms coordinate lists using the
transformation coefficients that are output by \prog{grmatch}. The
\prog{grtrans} code is also capable of fitting a general polynomial
transformation between point-pair lists if they are paired or matched
manually or by an external software. We should note that in the 
case of degeneracy, e.g.~when all points are on a perfect lattice,
the match will fail.

Both programs are part of the \fihat/HATpipe package that is under
development for the massive data reduction of the HATNet data-flow.
They can be easily embedded into UNIX environments, as both of them
parses wide-range of command line arguments for defining the structure
of the input data and fine-tuning the algorithm. The programs are also
capable of redirecting their input and/or output to standard streams.

By combining \prog{grmatch} and \prog{grtrans}, one can easily derive
the World Coordinate System (WCS) information for a FITS data file. 
Output of WCS keywords is now fully implemented in \prog{grtrans},
following the conventions of the package 
\prog{WCSTools}\footnote{http://tdc-www.harvard.edu/wcstools/}
\citep[see][]{mink02}.
Such information is very useful for manual analysis with well-known FITS
viewers \citep[e.g.~\prog{ds9}, see][]{joye03}. For a more detailed
description of WCS see \citet{calabretta02} and on the representation
of distortions see \citet{shupe05}.

The package containing the programs \prog{grmatch} and \prog{grtrans}
and other related software are accessible after registration
from the web address \texttt{http://www.hatnet.hu/software}.

\subsection{Performance on large data sets}
\label{sec:perf}

We used \prog{grmatch} and \prog{grtrans} to perform astrometry and
star identification on a large set of images taken by the HAT Network
of telescopes \citep{bakos04}.  The results presented in this paper are
based on observations originating from the following HATNet telescopes:
HAT-5, HAT-6 and HAT-7 located at the Fred Lawrence Whipple Observatory
(FLWO), Arizona\footnote{HAT-10 is also located at FLWO, but its data
was not used in this paper.}, plus HAT-8 and HAT-9 at the Smithsonian
Submillimeter Array roof (SMA) atop Mauna Kea, Hawaii. To recall, these
telescopes have an identical setup: $f=200\mathrm{mm}$,
$d=f/1.8$ telephoto lens and a \ccdsize{2K} CCD yielding an
$8\arcdeg\times8\arcdeg$ FOV. 
In order to test the method on different instruments, we also performed
astrometry on data taken by the TopHAT (FLWO) photometry follow-up
instrument. TopHAT is a 0.26m diameter, f/5 Ritchey-Cr\'etien design
with a Baker wide-field corrector, aided by a \ccdsize{2K} Marconi
chip, yielding 1.3\arcdeg\ FOV.

The steps of the astrometry and identification were the following.
First, for all observed fields, reference star lists were generated
using the 2MASS catalog \citep[see][]{skrutskie06} as reference.  These
reference lists include the source identifiers, the original celestial
coordinates (RA, Dec), an estimated I-band magnitude and the
$(\xi,\eta)$ projected coordinates of the stars.  We used arc
projection \citep[see][]{calabretta02} centered at the nominal center
of the given field, and scaling of the projection was unity in the
manner that a star located at 1 degree distance from the center of the
given celestial field has a unit distance in the $(\xi,\eta)$ plane
from the origin in the reference list. The FOV of the reference lists
were a bit wider than the nominal FOV of the HAT telescopes so as to
ensure a full overlap between the two lists in spite of the small
uncertainties in the positioning of the telescopes.

Second, an input star list was generated for each image, using our star
detection algorithm \prog{fistar}\ (also part of \fihat/HATpipe) that
detects and fits star-like objects above a given S/N threshold. This
detection yields a set of input lists that include the pixel
coordinates, $(X,Y)$ of the stars and other quantities (including the
flux, FWHM and the shape parameters).

Third, for each image, the input star list and the relevant reference
star list was matched using the program \prog{grmatch}. The match was
performed between the projected reference coordinates,
$(\xi,\eta)$, and the detected pixel coordinates, $(X,Y)$. The program
outputs two files: the list of the matched lines (this is the ``match''
file) and a small file that includes the fitted polynomial
transformation parameters and some statistical data (this is the
``transformation'' file). It should be emphasized that the match was
not done directly using the original celestial coordinates, as they
exhibit an unwanted curvature in the field.

Finally, the reference star list $(\xi,\eta)$ was transformed by the
program \prog{grtrans} into the system of the image $(X,Y)$ using the
``transformation'' file.  The transformed list shows where each star
with a given identifier would fall on the image. The transformation can
be also used to calculate the WCS information for the given image.

\begin{figure}
\plotone{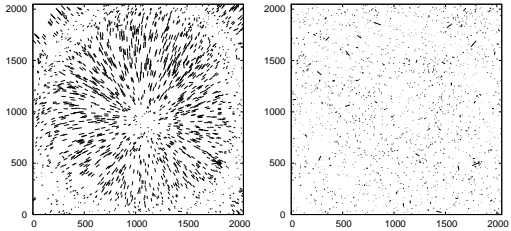}
\caption{
Vector plots of the difference between the transformed reference and
the input star coordinates for a typical HAT field. The left panel shows
the difference for second-order, the right panel for fourth-order
polynomial fits.
\label{fig:distvect}}
\end{figure}

\notetoeditor{This is the original place of \figr{distvect}. This is a
two-panel figure. We would request placing this figure close to the
context, and in such a way that one panel width is roughly one column
in the paper.}

\begin{figure}
\plotone{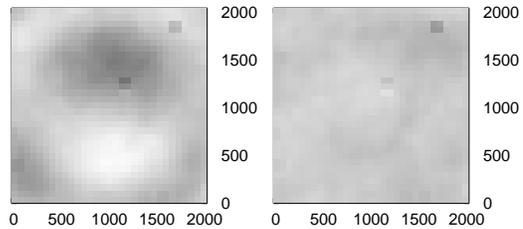}
\caption{
The difference between the $Y$ coordinates of the transformed reference
and the input star coordinates for a typical HAT field. The left panel
shows the difference for fourth-order, the right panel for sixth-order
polynomial fits.
\label{fig:distshift}}
\end{figure}

We note that the crucial part of the process is the third step. This
can be fine-tuned by many parameters, one of the most important being
the polynomial order. For a small FOV (less than one degree) and small
distortions, linear or second order polynomials yield good result. For
HAT images, we had to increase the order up to 6 to achieve the best
results. \figr{distvect} exhibits two vector plots that show the
difference between the transformed reference coordinates and the
detected star coordinates using a second and a fourth-order polynomial
transformation for a typical HAT image. In the first case, using a
second-order fit, definite radial structures remain, and the stars
located at the corners of the image are not even matched due to the
large distortions in the optics. Using a fourth-order fit, however, all
segments of the image are matched and the residuals are also smaller.
These small residuals can be better visualized if only the difference
between one of the coordinates is shown in a gradient plot:
\figr{distshift} illustrates the difference between the $Y$ coordinates
for the same image using a fourth- and sixth-order polynomial fit.
While there is a definite residual structure in the fourth-order fit,
it disappears using the sixth-order polynomial transformation.

\notetoeditor{This is the original place of \figr{distshift}. This is a
two-panel figure. We would request placing this figure close to the
context, and in such a way that one panel width is roughly one column
in the paper.}

As regards statistics, we performed the astrometry and source
identification for 243,447 HAT images that had been acquired between
the beginning of 2003 and June 2006. The wide-field telescopes observed
52 individual and almost non-overlapping fields between
$b=-30^\circ$ and $b=+74^\circ$ galactic latitudes. 

We initiated the processing with the following parameters. For the
triangulation, the 3000 brightest sources were used from both the
reference catalog and the detected stars. The critical unitarity was
set to $0.01$, therefore if the fitted initial transformation had an
unitarity larger than this value, the level of the triangulation
expansion was increased. The final transformation was determined using
a weighted sixth-order polynomial fit. Because the astrometric errors
of brighter stars are smaller, we weighted data-points based on their
magnitude during the fit. Finally, the maximal distance of matches was
set to one pixel to reject false identifications.

Astrometry and cross-identification of sources was successful for
238,353 images. The remaining 5094 images were analyzed manually, and
we found that only 13 of them were good enough to expect astrometry to
succeed; all the other ones were cloudy or had shown various other
errors. Astrometry on these 13 images also succeeded by decreasing the
number of stars for triangulation to 2000. It means that a completely
automatic run yielded 99.995\% success rate, and the other images were
also matched by applying small changes to the fine-tune parameters.

In order to test the algorithm with a different instrument, we also
performed astrometry on 22,936 TopHAT images taken in 2005. The only
difference in the procedure was that the polynomial transformation was
only of 2nd order. The success ratio was 93\%, but 90\% of the frames
where astrometry failed were cloudy with virtually no stars. Astrometry
also failed on very short exposure (10sec) V-band frames. Fine-tuning
the parameters (number of triangles, input lists) resolved most of
these cases.

The following statistics was done on the wide-field HAT frames. The
median value of the number of matched sources relative to the number of
stars in the reference or the input list was $98.38\%\pm0.31\%$ (median
deviation). The average value of the CPU usage was $0.77\pm0.22$
seconds per frame on a 64-bit AMD Opteron machine running at 2GHz.
Astrometry was successful on 96.78\% of the images using
Delaunay-triangulation without extended triangles (CPU: 0.73sec), while
0.49\% of the frames were processed at level $\ell=1$ extended
triangulation (CPU: 1.79 sec), 0.06\% at $\ell=2$ (CPU: 2.22 sec), 33
images at $\ell=3$ (CPU: 3.67 sec) and 1334+5094 images at $\ell=4$
extended triangulation (CPU: 5.20 sec). Here 5094 refers to those
images were astrometry failed even at $\ell = 4$, mostly because of bad
data quality (see before).
The reason for the Delaunay triangulation being successful for 96\% of
the wide-field HAT frames {\em without} extended triangulation is
because the HAT instruments perform homogeneous data acquisition, and
are very well characterized (zero-points, saturation). Thus, the 2MASS
reference catalogs can be retrieved for the given field in such a way
that there are many sources in common. In general applications,
however, when the saturation limit and faint magnitude limits of an
image have only a crude estimate, the extended triangulation is
essential.

Although the procedure is fast, we note that the most time consuming
part of the process is the triangulation generation and the triangle
matching itself.  On average, this required more than 60\% of the total
time, and at $\ell=4$, 92\% of the time.
The median value of the fit residuals was 0.06 pixels, while the median
value of the unitarities was 0.0042. The latter is in a quite good
agreement with the expected value of the nonlinearity factor
$f_d\approx0.005$.

%
\subsection{Comparison with other implementations}

We also compared the performance of the program \prog{grmatch} with an
existing implementation within IRAF, namely the \prog{images.immatch}
package with its relevant tasks \prog{xyxymatch}, \prog{geomap} and
\prog{geoxytran}. The steps of the point matching were as
described in \secr{pract}. First, an initial set of possible pairs were
established using \prog{xyxymatch} and the ``triangles'' option as
matching method. Because the triangle sets generated by
\prog{xyxymatch} are full triangulations, we limited our input lists to
the brightest sources, otherwise the $\ordo(N^3)$ dependence of the
number of the triangles would have resulted in an unrealistically long
matching time. Second, the initial transformation was fitted using
\prog{geomap}, and followed by transforming the reference catalog to
the frame of the input list using \prog{geoxytran} and this fit. Third,
the transformed reference and the original input list were also
matched by \prog{xyxymatch}, but this time using the ``tolerance''
matching method. Finally, this new list of point pairs were used again
to refine the geometrical transformation by \prog{geomap}.

The comparison of \prog{grmatch} and the IRAF \prog{images.immatch}
implementation was based on 950 individual images, all acquired by
TopHAT from the same FOV. We note that we had to use the relatively
narrow field TopHAT for the comparison, as the triangle match on the
original 8.2\arcdeg\ HATNet frames is almost hopeless given the spatial
distortions, the large number of stars, and the difficulty to select
the brightest stars and at the same time retain a small total number of
selected sources (in order to be able to cope with a full
triangulation). On each image there were approximately $800-900$
detected stars, depending on the airmass or thin clouds. For the
triangulation and the initial \prog{xyxymatch} fit we used the $35$
brightest sources both from the reference catalog and from the input
star lists. We found that \prog{grmatch} required $\sim 0.1$~sec CPU
time on average while the whole procedure using these IRAF-based tasks,
as described above, required $\sim 5-7$~sec net CPU time for a single image.
Both algorithms yielded the same transformation coefficients and
found the same number of pairs, however, in $3$ cases, the number
of sources used for triangulation had to be increased manually to 
$40$ or $45$. It is noteworthy that although the IRAF version proved to
be significantly slower, the time consuming part was the first
\prog{xyxymatch} matching. All other tasks, including the second
matching (with ``tolerance'' option) required only a fraction of a
second per image.


\section{Summary}
\label{sec:sum}

In this paper we present a robust algorithm for cross-matching two
two-dimensional point lists. The task is twofold: finding the smooth
spatial transformation between the lists, and cross-matching points. 
These two steps are intertwined, and are performed in an iterative way
till satisfactory transformation and matching rate are reached. We
make only very basic assumptions that hold for almost all astronomical
applications, including wide-field surveys with distorted fields and
large number of sources. Namely, the transformation between the point
lists is dominantly a similarity transformation (arbitrary shift,
rotation, magnification, inversion). A significant distortion term can
be present, given it can be linearized on the scale-length of the
average distance of neighboring points.

In \secr{pmatch} we briefly described symmetric point-matching in one
and two dimensions, because this tool is used throughout the astrometry
procedure. Finding the initial transformation between the point lists
is based on triangle matching. First we defined various normalized
triangle spaces in \secr{trispace}. The ``mixed'' triangle space of
\citet{valdes95} is a continuous function of the triangle parameters,
but flipped triangles are not distinguished. The ``chiral'' triangle
space ensures that chirality information is preserved, but this space
is not continuous. We showed that it is possible to define a
``continuous'' triangle space that is both continuous and preserves
chirality, and which, furthermore, spans a larger volume and diminishes
confusion of triangles having similar coordinates. 

Taking into account the distortion of a field and the astrometric
errors, we calculated both the optimal size and number of triangles.
For the typical setup of a HATNet telescope ($8\arcdeg\times8\arcdeg$
FOV, distortion factor $f_d \sim 0.005$) the optimal size is
$0.2\arcdeg$, and the optimal number of triangles is less than $2\times
10^6$. We use Delaunay triangulation for generating the triangles of
the triangle-space. This has the advantage of being fast, robust, and
generating local triangles that are less prone to being distorted. In
\secr{extdel} we noted, however, that Delaunay triangulation is
sensitive for removing or adding points to the list, and thus instable.
We introduced an extension of this triangulation that is
parametrized by an $\ell$ level.

Having determined the transformation between the two lists, one can
check how well our initial assumption of the dominant term being linear
holds. In \secr{unit} we introduced the unitarity of the
transformation, a simple scalar measure of this property. We described
the practical details of the algorithm in \secr{pract}, and the actual
software implementation (\prog{grmatch}, \prog{grtrans}) in \secr{sw}.

Finally, we ran the above programs on some 240,000 frames taken by the
wide-angle cameras of HATNet, plus 20,000 frames acquired by the TopHAT
telescope. The success rate was very close to 100\%, and the routines
handled the various pointing errors, defocusing and 6th order
distortions in the wide fields. Both programs will become available in
binary format for a wide range of architectures upon request from the
authors.


\acknowledgements

A.~P.~would like to thank the hospitality of the Harvard-Smithsonian
Center for Astrophysics, where this work has been partially carried
out. A.~P.~was also supported by the Hungarian OTKA grant T-038437.
The HATNet project is funded by NASA grant NNG04GN74G. 
G.~\'A.~B. wishes to acknowledge funding by the NASA Hubble Fellowship grant
HST-HF-01170.01-A. Both authors would like to thank Istv\'an Domsa for
the early development of triangle and point-matching codes.


{}


\end{document}